# DRR-MDPF: A Queue Management Strategy Based on Dynamic Resource Allocation and Markov Decision Process in Named Data Networking (NDN)

Fatemeh Roshanzadeh[1,2], Hamid Barati*[1,2], Ali Barati[1,2]

roshanzadeh@iau.ac.ir, hamid.barati@iau.ac.ir, alibarati@iau.ac.ir

1. Department of Computer Engineering, Dezful Branch, Islamic Azad University, Dezful, Iran
2. Institute of Artificial Intelligence and Social and Advanced Technologies, Dez.C., Islamic Azad University, Dezful, Iran

**Abstract:** Named Data Networking (NDN) represents a transformative shift in network architecture, prioritizing content names over host addresses to enhance data dissemination. Efficient queue and resource management are critical to NDN performance, especially under dynamic and high-traffic conditions. This paper introduces DRR-MDPF, a novel hybrid strategy that integrates the Markov Decision Process Forwarding (MDPF) model with the Deficit Round Robin (DRR) algorithm. MDPF enables routers to intelligently predict optimal forwarding decisions based on key metrics such as bandwidth, delay, and the number of unsatisfied Interests, while DRR ensures fair and adaptive bandwidth allocation among competing data flows. The proposed method models each router as a learning agent capable of adjusting its strategies through continuous feedback and probabilistic updates. Simulation results using ndnSIM demonstrate that DRR-MDPF significantly outperforms state-of-the-art strategies including SAF, RFA, SMDPF, and LA-MDPF across various metrics such as throughput, Interest Satisfaction Rate (ISR), packet drop rate, content retrieval time, and load balancing. Notably, DRR-MDPF maintains robustness under limited cache sizes and heavy traffic, offering enhanced adaptability and lower computational complexity due to its single-path routing design. Furthermore, its multi-metric decision-making capability enables more accurate interface selection, leading to optimized network performance. Overall, DRR-MDPF serves as an intelligent, adaptive, and scalable queue management solution for NDN, effectively addressing core challenges such as resource allocation, congestion control, and route optimization in dynamic networking environments.

**Key words:** Named Data Networking (NDN), Queue Management, Markov Decision Process Forwarding (MDPF), Deficit Round Robin (DRR), Adaptive Resource Allocation, Markov Decision Process (MDP).

## 1. Introduction

With the rapid growth of the Internet and the increasing number of users and applications, numerous challenges have emerged within the current core Internet architecture. These challenges include inefficiencies in adapting to new applications and evolving Internet requirements [1,2]. While Internet users are primarily concerned with accessing the desired content rather than its storage location, the research community is actively working toward designing a new architecture with fewer limitations. This new architecture is content-centric rather than host-centric. One of the most advanced and promising architectures in this area is Named Data Networking (NDN). NDN introduces a name-based communication model in which each piece of content is identified by a unique name and can be cached in various routers across the Internet. This feature enables data to be retrieved without dependence on a specific location. In NDN, data transmission is initiated by consumers sending Interest packets, which are then responded to with Data packets from the data producers. Routers forward Interest packets based on the data name while maintaining information about pending requests. This process enables routers to detect loops, evaluate the performance of different paths, and quickly identify and test alternative routes in case of failures. Routing strategies in NDN play a crucial role in optimizing network performance and can significantly affect the utilization of multi-connectivity capabilities and the overall efficiency of the network [3].

In Named Data Networking (NDN), the process of forwarding Interest packets faces several challenges. These challenges primarily stem from the unique characteristics of NDN, such as in-network caching, name-based routing, and the management of data requests initiated by consumers. One of the key issues in such networks is the synchronization and matching of Interest packets with corresponding data resources. This problem, often due to incomplete or imprecise name matching, can lead to increased delays in data delivery. Research has shown that the process of routing Interest packets using the Pending Interest Table (PIT) may encounter issues such as network congestion and suboptimal routing paths. These issues result in decreased network efficiency and increased data retrieval delays [4]. As presented in Figure 1, the Interest packet forwarding process in the NDN architecture involves the use of the PIT to store pending requests and the Forwarding Information Base (FIB) to select optimal paths toward data sources.

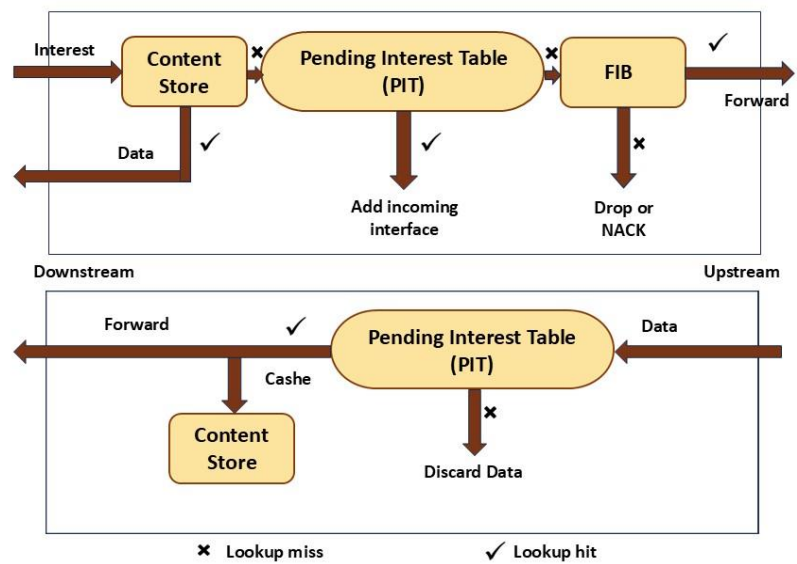

Figure. 1: Data forwarding process in an NDN node [2]

In addition, in-network data caching itself introduces challenges related to cache coordination and data consistency. Particularly when data is stored across multiple nodes, managing and updating caches to prevent resource misuse becomes one of the main challenges. If data is not properly cached or outdated and incorrect data is delivered to consumers, serious security and performance issues may arise [5].

Another challenge in the Interest forwarding process in NDN is the expiration of Interest packets. This problem occurs when the Interest timeout expires before receiving a response from data sources. In such cases, the corresponding entries in the Pending Interest Table (PIT) must be removed, and the requests need to be retransmitted, which can lead to increased network congestion and reduced efficiency in data delivery. Efficient strategies for managing timeout and optimizing the forwarding process can significantly improve NDN performance [4].

In this paper, an efficient and intelligent routing method for Named Data Networking (NDN) is introduced, aiming to improve resource utilization and enhance network performance. Routers in NDN must dynamically select the best outgoing interface for incoming requests, which depends on the current network state. Since route selection is a sequential decision-making problem, it is modeled as a decision process in stochastic environments.

Among various decision-making methods, the Markov Decision Process (MDP) has been chosen as a powerful framework to address such problems. Given that network conditions are influenced by the stochastic patterns of content requests and data paths, MDP is well-suited to make effective decisions in such random environments. This method has been specifically applied for flow allocation in heterogeneous access networks and facilitates improved performance in data networks.

In NDN, decision-making must naturally be capable of adapting to network changes. However, a key challenge in designing routing strategies is that, due to the transient and temporary nature of cached items, making routing decisions based on instantaneous changes in network parameters can lead to significant communication overhead. Therefore, employing MDP as an optimal decision-making method can address these challenges and effectively improve routing performance [6].

Considering the specific characteristics of the routing process in Named Data Networking (NDN), this paper proposes an intelligent and adaptive method for routing requests based on the integration of the Markov Decision Process (MDP) and the Deficit Round Robin (DRR) resource management algorithm. MDP, as an effective decision-making method under stochastic and uncertain network conditions, enables the selection of the best outgoing interface based on the network state.

The DRR algorithm, as a queue and resource management approach, helps the network efficiently manage resources and reduce delay by fairly and optimally allocating bandwidth. This combination of MDP and DRR is specifically designed for routing in NDN networks, aiming to improve network efficiency and optimize resource utilization under dynamic and varying conditions. The proposed method leverages the advantages of optimal decision-making in MDP along with the DRR algorithm to enable the network to adapt to instantaneous changes in network conditions and make the best use of resources.

In traditional queue management approaches, decisions are primarily made based on static rules, which leads to reduced network flexibility in coping with dynamic changes. In contrast, in the proposed method, queue management decisions are made not only based on historical data and computed rewards but also by considering the instantaneous network state. Thus, the choice of queuing method in this strategy does not rely on a single parameter but is a combination of analyses based on the Markov Decision Process (MDP) and the fair allocation mechanisms of Deficit Round Robin (DRR).

The goal of this method is to improve resource allocation decisions and enhance network efficiency through a continuous and adaptive learning model. In the proposed method, each router is considered as a decision-making agent that must adopt the optimal strategy for queue management. The Markov Decision Process is utilized to analyze the network state and predict future conditions, while DRR serves as a fair resource allocation mechanism to improve queuing performance. In this method, three key parameters are considered for queue management:

The proposed DRR-MDPF method has several fundamental differences from other methods presented in the literature such as:

- First, in terms of modeling tools and problem-solving techniques, it is the first method to combine Markov Decision Process-based Forwarding (MDPF) with Deficit Round

Robin (DRR). In this method, MDPF is used to optimize routing decisions based on rewards and past experiences, which leads to improved path selection and better performance under various network conditions.
- Second, unlike many existing methods that are multipath-based, the proposed method is a single-path method. While multipath techniques can distribute resources across several routes, DRR-MDPF has demonstrated superior performance in metrics such as load balancing and packet drop rate.
- Third, and most importantly, the number and type of metrics considered for optimal path selection are significantly broader. Most existing methods in Table 1 rely on only one or two metrics for route selection, whereas DRR-MDPF takes into account three critical factors: bandwidth, round-trip time (RTT), and the number of unsatisfied requests. These metrics contribute to improved network efficiency, Interest Satisfaction Rate (ISR), and content retrieval time.

The obtained results indicate that the proposed method demonstrates significant improvements over other methods across most performance evaluation metrics, including network load balancing, average throughput, packet drop rate, Interest Satisfaction Rate (ISR), and average content retrieval time.

The proposed method in this paper is the first to introduce the combination of MDPF and DRR strategies to enhance queue management performance in Named Data Networking (NDN). It acts as an intelligent and adaptive strategy capable of optimally managing network resources under various conditions. The main contributions of this work are summarized as follows:

- **Queue management based on Markov Decision Process (MDP) and Deficit Round Robin (DRR):** In this method, the router is modeled as a decision-making agent that uses MDP to predict and optimize packet forwarding decisions. DRR is employed as a fair resource allocation mechanism. This combination enhances queue performance and ensures optimal resource distribution.
- **Multi-metric path selection:** Unlike existing methods that typically consider only a single metric for path selection, the proposed method uses three criteria—bandwidth, round-trip time (RTT), and the number of unsatisfied requests—to make routing decisions. This results in more accurate path selection and reduced data delivery latency.
- **Adaptability to sudden network topology changes:** The proposed method is capable of intelligently handling unexpected changes in network topology. By leveraging feedback from the network, it selects optimal paths for forwarding requests, avoiding bottlenecks and link failures.
- **Enhanced network performance:** Simulation results using ndnSIM demonstrate that the proposed method outperforms existing approaches under heavy traffic loads and dynamic network conditions. Specifically, it shows significant advantages in resource allocation, delay reduction, load balancing, bandwidth efficiency, and packet loss minimization.

The first section presents an introduction to the challenges of queue management in Named Data Networking (NDN) and highlights the necessity of employing hybrid method to enhance performance. The second section reviews related work in the areas of resource allocation, queuing algorithms, and Markov decision-making methods. The third section covers the fundamental concepts. The fourth section is dedicated to a comprehensive introduction of the proposed DRR-MDPF algorithm, including the description of its modules, the learning process, and the integration of DRR with MDPF. In the fifth section, the implementation of the algorithm within the ndnSIM simulation environment is described, along with the performance evaluation metrics. The sixth section presents and analyzes the simulation results, comparing the performance of DRR-MDPF with existing algorithms. Finally, the seventh section offers a conclusion, summarizes the key findings, and suggests directions for future research.

## 2. Related Works

Boyan and Littman (1993) proposed a reinforcement learning-based method for packet routing in dynamically structured networks. This method, by leveraging adaptive learning in response to environmental changes, enhances routing decisions based on the current network state. While the proposed algorithm performs well in environments with frequently changing topologies, it requires longer training periods to converge to optimal policies and may suffer from instability under rapid changes [7].

Wang et al. (2013) introduced an improved algorithm titled Hop-by-Hop Interest Shaper to manage congestion in Named Data Networking (NDN). This algorithm employs interest shaping techniques to efficiently control traffic flow at each routing step, thus preventing network congestion. Although effective in reducing delay and enhancing performance under high traffic, it may face stability and efficiency challenges in networks with highly variable traffic patterns [8].

Yi et al. (2014), in their paper On the Role of Routing in Named Data Networking, examined the importance of routing optimization in NDN. They presented various strategies for optimal path selection and delay reduction, emphasizing the need for intelligent routing algorithms in networks with heavy traffic and complex topologies. However, they also acknowledged the challenges of implementing such algorithms at large scale [9].

Compagno et al. (2015), in the context of Information-Centric Networking (ICN), explored the role of negative acknowledgments (NACKs) in the routing process. They proposed an algorithm utilizing NACK messages to enhance detection of failed paths and optimize resource management. While this method can reduce delays in request retransmissions and improve stability during congestion, designing efficient NACK messages without introducing significant overhead remains a key challenge [10].

Pusch et al. (2016) introduced an algorithm called Stochastic Adaptive Forwarding (SAF) that uses stochastic processes to dynamically select optimal paths for data forwarding in NDN. While effective under dynamic network conditions, its path selection accuracy may degrade in more complex scenarios [11].

Croffilio et al. (2016) proposed the Request Forwarding Algorithm (RFA), which combines routing requests to optimize request forwarding in NDN. Although this algorithm improves network performance under high traffic, excessive request volumes can introduce delays and reduce overall network efficiency [12].

Chen et al. (2017) presented Fuzzy Interest Forwarding, an algorithm based on fuzzy logic to adaptively manage interest forwarding in NDN. Designed for uncertain or incomplete network information, this approach is particularly useful in environments with high traffic or topological volatility. However, the complexity of rule design and inference engine implementation poses scalability challenges [13].

Zhang et al. (2017) proposed a name-based data synchronization method in NDN, aimed at improving scalability and efficiency in data exchange across distributed nodes. By using efficient naming structures and lightweight difference detection mechanisms, their method enables fast and reliable synchronization. Nevertheless, it may struggle to maintain stability and manage overhead under high update rates or resource-constrained conditions [14].

Yao et al. (2018) developed the Stochastic Markov Decision Process Forwarding (SMDPF) algorithm to optimize routing in NDN using a semi-Markov decision framework. While capable of making optimal decisions in dynamic environments, its high computational complexity may hinder scalability in large-scale deployments [15].

Ali and Lim (2021) proposed NameCent, a name-centric method for reducing data dissemination overhead in vehicular NDN (VNDN). By analyzing name structures and relative node positions, this method enables smarter data distribution decisions, reducing congestion and improving bandwidth utilization. Simulation results demonstrate its superior performance in delay reduction and traffic load management, especially in highly mobile and complex naming environments [16].

Abdi et al. (2022) introduced Learning Automata-based MDPF (LA-MDPF), which combines learning automata with Markov decision processes to dynamically adjust routing in NDN. This algorithm is effective in dynamic networks with variable traffic loads. However, its inherent complexity and computational demands make real-world implementation challenging [17].

Pan et al. (2022) proposed a novel active queue management algorithm based on monitoring the rate of change in average queue length. Unlike traditional methods that rely solely on current queue length, this algorithm reacts more promptly to traffic fluctuations, resulting in reduced delay and improved throughput. Simulation results show it outperforms classic AQM techniques like RED and PI [18].

Khan and Lim (2024) introduced a real-time vehicle tracking-based data forwarding method in VNDN using the Recursive Least Squares (RLS) algorithm. By dynamically analyzing vehicle positions, the method enables more accurate path prediction and efficient relay node selection, achieving lower delay and higher delivery ratio in high-mobility environments [19].

Ma et al. (2024) proposed a hybrid congestion control scheme for NDN leveraging Software-Defined Networking (SDN) capabilities. With centralized real-time monitoring of nodes and traffic, the SDN controller optimizes flow management and resource allocation. This integrated SDN-NDN method shows improved congestion reduction and service quality in high-traffic scenarios [20].

Lefevre and Yaghoubi (2024) presented an analytical study on optimal control in queueing systems, using control theory to reduce waiting time and improve system efficiency. By precisely formulating the problem and applying optimization methods, they demonstrated how adjusting arrival and departure rates can enhance performance under varying load conditions. This work lays a theoretical foundation for designing efficient queue management algorithms in complex networks [21].

Rodriguez-Perez et al. (2024) proposed an end-to-end active queue management method for NDN to improve data transmission across full communication paths. This model applies coordinated queue control not only at intermediate nodes but also throughout the entire route, combining local and global network perspectives to make more informed packet retention and drop decisions. Experimental results indicate significant improvements in delay reduction and data delivery success under heavy loads [22].

Matroud and Ali (2025) conducted a comparative study of traditional and AI-enhanced active queue management (AQM) techniques for congestion control. By evaluating algorithms like RED, CoDel, and their intelligent variants using machine learning and neural networks, they showed that AI-based methods offer superior prediction accuracy and responsiveness to traffic changes. Their findings highlight the enhanced adaptability and reduced latency of smart AQM in dynamic network environments [23].

A comparison of the methods listed in Table 1 is provided.

Table 1: A Comparative Overview of Various Methods in Named Data Networking (NDN)

| Method | Key Contribution | Advantages | Limitations |
|---|---|---|---|
| Q-Routing [7] | Introduced reinforcement learning for routing in dynamic networks | High adaptability to network changes | Slow convergence and high resource usage in large networks |
| OFTT [8] | Hop-by-hop interest shaping for congestion control in NDN | Reduced delay during congestion | Requires fine-tuned parameters for various topologies |
| NCC [9] | Analyzed routing role in NDN performance | Clarified key routing components | Limited actionable solutions provided |
| iNRR [10] | Explored NACK impact for feedback and congestion control | Improved error recovery capability | Possible overhead from negative feedback |
| SAF [11] | Stochastic adaptive forwarding mechanism in NDN | Better load distribution via path diversity | Depends on probability tuning and slow learning |
| RFA [10] | Optimal multipath congestion control and smart forwarding | Enhanced network efficiency | Complex implementation |
| Fuzzy Forwarding [11] | Fuzzy logic-based forwarding for multi-metric decisions | Adaptability in uncertain environments | Hard to configure fuzzy rules |
| PIF [12] | Scalable name-based data synchronization | High scalability | Less effective in dynamic scenarios |
| SMDPF [13] | Forwarding based on Semi-Markov Decision Process | State-aware smart decision making | Requires complex modeling |
| NameCent [14] | Name centrality to reduce broadcast in vehicular NDN | Reduced control overhead | Less accurate in non-vehicular networks |
| LA-MDPF [15] | Combines learning automata and MDP for optimal forwarding | Strong in dynamic conditions | High complexity and resource usage |
| AQM-AQLCR [16] | AQM using average queue length change rate | Fast reaction to traffic changes | Sensitive tuning to avoid oscillation |
| RLS-based Forwarding [17] | Real-time vehicle tracking for NDN forwarding | Higher routing accuracy and lower delay | High computational overhead in dense networks |
| SDN Hybrid Control [18] | Hybrid SDN-based congestion control for NDN | Global view, quick congestion response | Requires SDN infrastructure and coordination |
| Optimal Queue Control [19] | Optimized delay reduction in queuing systems | Accurate mathematical decision-making | Mostly theoretical, less real-time practicality |
| End-to-End AQM [20] | End-to-end AQM for NDN | Improved QoS, lower congestion | Complex implementation, needs inter-node synchronization |
| AI-AQM Comparison [21] | Comparison of traditional vs. AI-based AQM | High effectiveness and traffic adaptability | Dependent on training data and AI models |

## 3. Fundamental Concepts

In the following, the fundamental concepts and algorithms employed in the proposed method for optimal resource allocation and performance enhancement in Named Data Networking (NDN) architecture are introduced.

### 3.1 Deficit Round Robin (DRR) Algorithm

The Deficit Round Robin (DRR) algorithm effectively allocates resources within networks and ensures high reliability in traffic management. DRR adopts a mechanism similar to Round Robin (RR) for distributing resources among flows but dynamically adjusts the packet transmission permissions based on system status and demand. This adaptive behavior enables DRR to outperform traditional approaches under variable network load conditions. According to Jain et al. (1984), resource allocation in shared systems can be managed efficiently in terms of both fairness and performance, and algorithms such as DRR have been proven to perform well in this regard [22].

### 3.2 Markov Decision Process (MDP)

Named Data Networking (NDN), as one of the emerging models in modern communication networks, requires optimal strategies for request forwarding and queue management. One of the effective method in this context is the application of Markov Decision Process (MDP), combined with various strategies to improve resource allocation and decision-making under diverse network conditions.

MDP is recognized as a mathematical framework for decision-making in stochastic and fully observable environments [28]. A Markov Decision Process consists of a set of states (S), actions (A), rewards (R), and a state transition probability matrix (P), which are updated sequentially over time. The primary objective is to select the best possible action at each time step to optimize the system's overall performance.

The Markov Decision Process can be formally modeled under various scenarios as described by Equation (1):

$$R(S(t)) = \sum_{S' \in S(t+1)} R(S(t), a, S'(t+1)) \, p(S'(t+1) \mid S(t), a) \tag{1}$$

Where $R(S(t))$ denotes the reward received at time t, The term $p(S'(t+1) \mid S(t)a)$ represents the transition probability from state $S(t)$ to state $S'(t+1)$ as a result of taking action a.

It is also assumed that the transition probabilities are normalized according to Equation (2):

$$R(S(t)) = \sum_{S' \in S(t+1)} p(S'(t+1) \mid S(t), a) = 1 \tag{2}$$

## 4. System Model

In this paper, we examine and analyze the combination of Markov Decision Process Forwarding (MDPF) and Deficit Round Robin (DRR) routing strategies to improve queue management performance in Named Data Networking (NDN). NDN, as a new network architecture, is based on sending Interest packets and receiving Data packets from various sources. Unlike traditional IP networks, which address devices, NDN operates based on content names. One of the main challenges in these networks is queue management in routers, which directly affects latency and routing efficiency.

In the proposed method, the network is modeled as a directed graph G = (N, L), where N is the set of routers and L is the set of communication links between them. This model allows routers to direct Interest packets toward the best data sources. The use of MDP and DRR strategies enables the network to intelligently leverage past experiences to make optimal decisions for queue management. In this combination, MDP, using Markov learning processes, provides optimal choices for routing requests, while DRR employs a balanced resource allocation approach for queue management.

The complete process of the proposed DRR-MDPF method is presented in Figure 2. In this method, two key modules, MDPF and DRR, are implemented jointly within NDN nodes to optimize the Interest forwarding process.

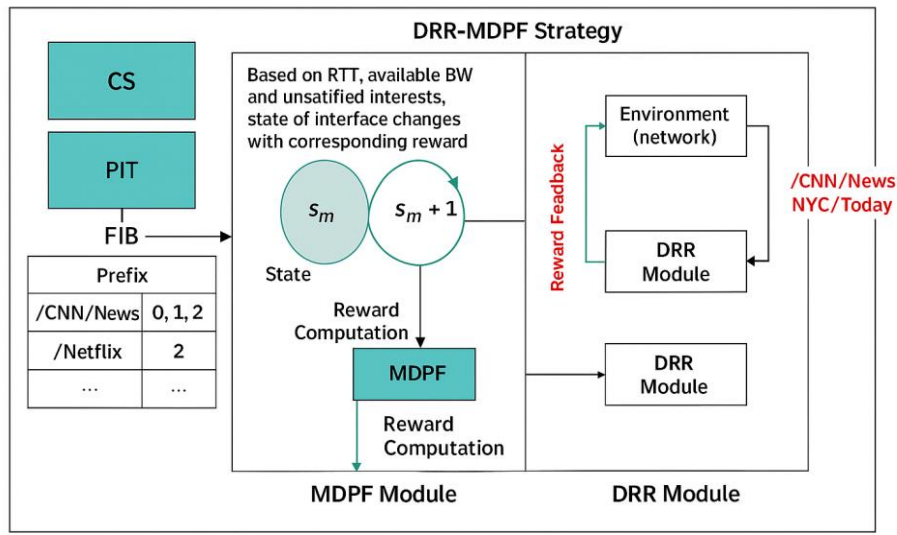

Figure 2. Block diagram of the proposed method

One of the key features of this model is in-network caching and the use of optimized caches within routers. These caches help routers store previous requests and prevent the retransmission of similar requests. Consequently, the combination of these two strategies not only improves routing performance but also significantly reduces network congestion and delays in data delivery. To clarify the conceptual model and the presented formulations, the symbols and parameters used in this paper are summarized in Table 2.

**Table 2:** Notations.

| Notation | Description |
|---|---|
| $a_i(t)$ | Action $a_i$ at time t (the output of automaton) containing the selected interface |
| $b_i(t)$ | Input (the response to action $a_i$) given to the automaton at time t |
| $q(t)$ | State of the automaton at time t |
| $n, m$ | Numbers of automaton's actions and states, respectively |
| $p_i$ | Probability of action $a_i$ |
| $c_i$ | Probability of penalty for action $a_i$ |
| $\lambda_r$ | Learning parameter ($0 < \lambda_r < 1$) |
| $S, A, R, P, T$ | Components of MDP: set of states, set of actions, set of rewards, probability matrix of transitions, and set of time points, respectively |
| $k$ | Content class |
| $l$ | Interface |
| $L$ | Total number of interfaces |
| $b_l, B_l$ | Available bandwidth and normalized bandwidth on interface $l$ |
| $c_{lk}, \theta_{lk}$ | Number of unsatisfied Interests and normalized number |
|   | of unsatisfied Interests for content class $k$ on interface $l$ |
| $d_{lk}, \delta_{lk}$ | Average delay and normalized delay for content class $k$ on interface $l$ |
| $S_l$ | State of interface $l$ |
| $R_L$ | Reward for interface $l$ |
| $Pro_l$ | Probability of selecting interface $l$ |
| $wpro_l$ | Weighted probability for interface |

## 5. Proposed Method

In the proposed method, the router is modeled as an intelligent agent in data-centric networks, utilizing a Markov Decision Process Forwarding (MDPF) mechanism to guide packet forwarding. This approach enables dynamic and informed decision-making based on the real-time state of queues, packet priorities, and network conditions, allowing for the optimal selection of both the path and timing for packet transmission.

Alongside this decision-making framework, the Deficit Round Robin (DRR) algorithm is employed as an efficient mechanism for fair resource allocation among data flows. By using dynamic quota distribution, DRR assists the router in evenly distributing bandwidth across queues and prevents bottlenecks on specific paths.

The combination of these two mechanisms in the DRR-MDPF approach leads to adaptive and intelligent decisions in packet forwarding. This solution results in reduced latency, improved throughput, and enhanced network resource efficiency, particularly under high-traffic and dynamic conditions in NDN environments.

### 5.1 Modeling the DRR Algorithm within the Framework of the Proposed Method

The Deficit Round Robin (DRR) algorithm, as a decision-maker in network systems, can effectively allocate resources. To model DRR, the algorithm can be considered as an automaton that dynamically assigns packets based on system conditions and demand. In this model, a set of actions (A) is defined as the operations performed by the automaton, including packet allocation to flows and queue management. A set of inputs (B) represents the various network conditions and states. Similarly, a set of states (Q) is considered for the automaton, reflecting different queue conditions and resource allocation statuses.

In this model, the transition function (F) updates the states and makes new decisions based on environmental feedback such as network load and delay. The transition function is defined according to Equation (3):

$$q_{t+1} = F(q_t, b_t) \tag{3}$$

Where $q(t)$ is the state at time $t$, and $b(t)$ is the input at time $t$. The output of the automaton, which determines the DRR decisions, is specified by the output function *(G)*. The output function is defined by Equation (4):

$$a(t) = G(q(t), b(t)) \tag{4}$$

Where $a(t)$ is the action selected at time $t$. This algorithm makes new decisions at each time step based on past experiences and dynamically updates the action probabilities. The action probabilities are represented by Equation (5):

$$\sum_{i=1}^{n} p_{i(t)} = 1 \tag{5}$$

As mentioned in the literature, resource allocation in shared systems can be effectively managed in terms of fairness and efficiency, and algorithms like DRR perform well in this regard [22].

### 5.2 Environment

The environment represents the entity that responds to the decisions of the automaton. In the proposed method, the router is considered as the automaton and the network as the environment. Mathematically, the environment can be represented by the triplet {A, B, C}, where:

- $A = \{a_1, a_2, ..., a_n\}$ is the set of possible actions that the router can perform (such as selecting packets for transmission or dropping packets).
- $B = \{b_1, b_2, ..., b_n\}$ is the set of environment responses that indicate the result of each action (e.g., success or failure in sending a packet).
- $C = \{c_1, c_2, ..., c_n\}$ is the set of penalty probabilities representing the likelihood of failure or delay in the environment's responses.

Based on the nature of the set B, environments are divided into three models:

- Model P: In this model, a desirable or successful response is represented by the number 0, and an undesirable or failed response is represented by the number 1.
- Model Q: Uses a limited and discrete set of environment responses, which can include numerical values or different states.
- Model S: Uses real numbers in the interval [0,1], which can represent the percentage of success, delay, or other continuous variables in the environment.

The set C, which represents the penalty (failure) probabilities in environmental responses, is defined by equation (6):

$$P_{penalty}(a_i, t) = P(\text{"Failure"} \mid a_i(t)) \tag{6}$$

where $a_i(t)$ represents the execution of the *i*-th action at time *t*. In the environments, the penalty probability for each action is considered fixed, but this probability can be updated over time.
In the combined MDPF and DRR model, the router's decisions as an automaton directly impact the network environment. Similarly, some studies have shown that using LA- and MDPF-based strategies can help improve packet forwarding in Named Data Networks (NDN) [8]. Therefore, the probability of receiving an undesirable response (such as high delay or packet loss) depends on the decisions made by the router and the current state of the network. For example, under heavy network load conditions, the router may need to fairly allocate packets using the DRR strategy and select the best path for forwarding packets using MDPF [23].
Through continuous repetition of this process, actions that enhance network performance are progressively chosen. This method can also improve system performance in stochastic environments, enabling the network to effectively utilize available resources [24].

### 5.3. Optimization Phase

In this section, the process of selecting the optimal action by the router from a set of options is examined. In the combined MDPF and DRR model, the router makes decisions based on the network state and past experiences. The router uses a set of possible actions (such as selecting a path or forwarding packets to different queues) and continuously optimizes these choices depending on the network conditions. During the optimization process, if the router selects an appropriate action that improves network performance (for example, reducing delay or preventing packet loss), this action is reinforced for future selections. Conversely, if the chosen action leads to an undesirable outcome (such as high delay or packet loss), the router learns to consider this choice less in the future. Specifically, in the DRR strategy, the router makes decisions using a fair algorithm for packet allocation. Packet allocation in this algorithm is performed according to Equation (7):

$$\delta_i(t+1) = \begin{cases} \delta_i(t) - L_i & if\ \delta_i(t) >= L_i \\ \delta_i(t) + Q_i & otherwise \end{cases} \qquad (7)$$

Where $\delta_i(t)$, $L_i$, and $Q$ are respectively the deficit counter of flow $i$ at time $t$, the packet size of the flow, and the Quanta value (a fixed amount added to the deficit counter).

In the MDPF strategy, the router selects optimal paths for forwarding packets using the Markov Decision Process. This process is modeled according to Equation (8):

$$V(s) = \max_{a \epsilon A} \left( R(s, a) + \gamma \sum_{s'} P(s' \mid s, a) V(s') \right) \qquad (8)$$

Where $V(s)$ is the value of state $s$, $\gamma$ is the discount factor representing the impact of future decisions on current choices, $P(s' \mid s, a)$ is the probability of transitioning from state $s$ to $s'$ by taking action $a$, $A$ is the set of possible actions, and $R(s, a)$ is the reward received for performing action $a$ in state $s$. This process continuously leverages past experiences to improve network performance, helping to reduce delay and enhance efficiency [25].

### 5.4. Markov Decision Process (MDP)
In this section, the Markov Decision Process (MDP) and its application in improving queue management performance in Named Data Networks (NDN) are examined. In NDNs, the Markov Decision Process can serve as an important tool for selecting optimal paths for requests and allocating resources in queues. This decision-making system helps the router make optimal decisions for forwarding packets and managing queues based on the current network state and past experiences. Specifically, in the combined strategy of MDP and DRR, the Markov decision process enables routers to select optimal paths for packet forwarding based on rewards and costs [26][27].

In the combination of these two strategies, routers first use the MDP process to select optimal paths based on the network state and past experiences. Then, using the DRR algorithm, packets are allocated to queues fairly and efficiently. This method not only reduces network delay but also prevents packet loss and improves network performance. In the Markov decision process, the network state and action choices can be updated sequentially over time. For example, if choosing a specific path leads to reduced delay or increased throughput, this action is emphasized in future decisions, increasing the likelihood of selecting that path again. Conversely, if the chosen path results in undesirable conditions (such as increased delay or packet loss), the router learns to select that path less frequently in the future.

### 5.4.1. State Space
In the DRR-MDPF algorithm, the state space represents the various states that each network interface may be in. These states are determined based on features such as available bandwidth, the number of pending requests, and data transmission delays [10]. For each interface l in the router, the state space is expressed according to Equation (9):

$$S_{l,k(t)} = (b_l, c_k, d_{l,k}) \tag{9}$$

Where $b_l$ is the available bandwidth for interface $l$, $c_l$ is the number of pending (unanswered) requests on interface $l$, and $d_{l,k}$ is the delay in data transmission through the interface to node $k$. The overall network state at time $t$ is calculated according to Equation (10):

$$S(t) = \{S_{l,k(t)}\} \tag{10}$$

This state space helps the router to consider the precise condition of the network and the status of each interface at any given time, allowing it to make more optimal decisions for data forwarding.

### 5.4.2. Action Space

In this algorithm, an action refers to selecting one interface from the set of available interfaces. Suppose each router has $L$ interfaces. Then, the action space is defined as in Equation(11):

$$A = \{a_1, a_2, \ldots, a_L\} \tag{11}$$

At each time step, the algorithm must optimally decide which interface to select. This selection is made to maximize the obtained reward. The decision to select an interface is mathematically defined by Equation (12):

$$a_k^* = argmax\left(R(S_{l,k(t)})\right) \tag{12}$$

### 5.4.3. Reward

The reward for each action depends on the status of the interfaces. The goal is to select interfaces that have higher available bandwidth, lower delay, and fewer pending requests. The reward function is generally defined as follows:

To allow the parameters to be used simultaneously and in combination, they need to be normalized. For this purpose, the normalized parameters are calculated according to Equation (13):

$$\beta_l = \frac{b_l}{\sum_{l=1}^{L} b_l} \qquad \theta_{lk} = \frac{c_{lk}}{\sum_{l=1}^{L} c_{lk}} \qquad \delta_{lk} = \frac{d_{lk}}{\sum_{l=1}^{L} d_{lk}} \tag{13}$$

In these equations, $\delta$, $\theta$, and $\beta$ represent the normalized values of bandwidth (b), number of pending requests (c), and delay (d) for interface $l$, respectively. Then, the reward function is calculated according to Equation (14):

$$R(S_l(t)) = \delta_l + \beta_l \cdot \theta_l \tag{14}$$

### 5.4.4. Probabilistic Updates

Initially, the probability of selecting each interface is equally distributed, as shown in Equation (15):

$$Pro_{l,k}(0) = 1/L \qquad (15)$$

Then, a weighted probability for each interface is calculated according to Equation (16):

$$wpro_{\{l,k\}}(0) = \frac{R(S_{l,k}) * Pro_{l,k}(0)}{\sum_{l \in L} R(S_{l,k}) * Pro_{l,k}(0)} \qquad (16)$$

To avoid excessive sensitivity or slow adaptation, a learning parameter $\lambda$ is introduced according to Equation (17) to dynamically update the selection probabilities:

$$P_{l,k}(t+1) = (1-\lambda)P_{l,k}(t) + \lambda w_{\{pro\}}(s_l(t)) \qquad (17)$$

Where $\lambda$ (ranging between 0 and 1) controls the adaptation rate. A smaller $\lambda$ results in slower changes, while a larger $\lambda$ may cause instability in the algorithm.

---
**Algorithm 1: Selecting the Best Interface for Interest Packets in DRR-MDPF**
---
1: Input: Number of interfaces L, number of content classes K
2: Output: Optimal interface selection for each state
3: Initialization
4: t = 1
5: for each content class k do
6:   Initialize the probability of selecting interface l for class k:
7: end for
8: Running
9: while true do
10:   if an Interest packet of content class k arrives then
11:     Determine the state for each interface
12:     Compute reward for each interface based on its current state
13:     Calculate weighted probabilities by multiplying the initial probabilities with rewards and normalizing:

$$wpro_{\{l,k\}}(0) = \frac{R(S_{l,k}) * Pro_{l,k}}{\sum_{l \in L} R(S_{l,k}) * Pro_{l,k}}$$

14: else if a feedback packet of content class k arrives then
15: if positive response then
16: Update probabilities:

$$p_{l,k}(t+1) = 1 - \sum_{\{j \neq l\}} \lambda_r p_{j,k}(t)$$
$$p_{j,k}(t+1) = \lambda_r p_{\{j,k\}}(t)$$

17: else
18: Maintain previous probabilities:
19: end if
20: end if
21: t=t+1
22: end while

---

To better understand the proposed method, consider the following example:

Suppose a router has five interfaces L=5. After determining the states of the interfaces and calculating the corresponding reward, the results are as follows: R(S1) = 0.25, R(S2) = 0.2, R(S3) = 0.3, R(S4) = 0.2, R(S5) = 0.2. Given the initial probability of $Pro_l(0) = \frac{1}{5} = 0.2$ The initial probability for each interface is $Pro_l(0) = \frac{1}{5} = 0.2$. Therefore, the weighted probabilities for the interfaces are as follows:

$$wpro_1(0) = \frac{0.25 * 0.2}{0.25 * 0.2 + 0.2 * 0.2 + 0.3 * 0.2 + 0.2 * 0.2 + 0.2 * 0.2} = 0.217$$

$$wpro_2(0) = \frac{0.2 * 0.2}{0.25 * 0.2 + 0.2 * 0.2 + 0.3 * 0.2 + 0.2 * 0.2 + 0.2 * 0.2} = 0.173$$

$$wpro_3(0) = \frac{0.3 * 0.2}{0.25 * 0.2 + 0.2 * 0.2 + 0.3 * 0.2 + 0.2 * 0.2 + 0.2 * 0.2} = 0.260$$

$$wpro_4(0) = \frac{0.2 * 0.2}{0.25 * 0.2 + 0.2 * 0.2 + 0.3 * 0.2 + 0.2 * 0.2 + 0.2 * 0.2} = 0.173$$

$$wpro_5(0) = \frac{0.2 * 0.2}{0.25 * 0.2 + 0.2 * 0.2 + 0.3 * 0.2 + 0.2 * 0.2 + 0.2 * 0.2} = 0.173$$

It can be observed that the third interface ($l=3$) has the highest weighted probability. Therefore, this interface will be selected with the highest probability for sending the Interest packet. If a positive response is received from the network, such as the arrival of a Content Packet, this probability will be updated in subsequent steps to increase the chance of selecting this interface. Given the higher probability of action $a_3$, it is assumed that choosing it leads to a positive outcome, resulting in the network assigning a reward to it, which in turn increases the probability of selecting $a_3$ again in the next step. As previously mentioned, the value of $\lambda_r$ is chosen close to 1; thus, it is assumed that $\lambda r = 0.9$.

$$Pro_3(1) = 1 - \lambda \sum_{\{j \neq i\}} \lambda_r p_{j(0)}$$

$$1 - 0.9 * (0.217 + 0.173 + 0.173 + 0.173) = 0.3376$$

$$Pro_1(1) = 0.9 * 0.217 = 0.1953$$

$$Pro_2(1) = Pro_4(1) = Pro_5(1) = 0.9 * 0.173 = 0.1557$$

In this example, action a3 has a higher chance of being selected in the next step because it has received good responses in the past. However, the proposed algorithm may also select other actions to avoid getting stuck in local optima.

### 6. Performance Evaluation
In this section, the performance of the proposed DRR-MDPF method is simulated and compared with other routing strategies in Named Data Networking (NDN). The proposed method is compared against the SAF [9], RFA [10], SMDPF [13], and LA-MDPF [15] algorithms.

The main objective of this study is to compare the performance of the proposed DRR-MDPF algorithm with other queue management strategies under various network conditions. For this evaluation, the ndnSIM simulator, which is an NS-3 module for implementing the NDN protocol, has been used. The performance of DRR-MDPF is assessed under different network load conditions considering various performance metrics including throughput, request satisfaction ratio, packet drop rate, delivery delay, and load balancing. The simulation parameters are shown in Table 3.

Table 3: Simulation Parameters.

| Parameter | Value |
| --- | --- |
| Number of Nodes | 40 |
| Number of Links | 122 |
| Interest Packet Rate | 2000 to 4000 Interests per second |
| Cache Size | From 1% to 60% of total content |
| Queue Size | 100 packets |
| Compared Algorithms | SAF, RFA, SMDP, LA-MDPF, DRR-MDPF |
| Simulation Tool | ndnSIM on NS-3 |
| Simulation Duration | 150 seconds |
| Data Rate | 10 Mbps |
| Link Capacity | 10 Mbps |
| Link Delay | 10 ms |
| Queue Scheduling Algorithm | DRR (Deficit Round Robin) |
| Evaluation Metrics | Throughput, Interest Satisfaction Ratio, Drop Rate, Delay, Load Balancing |

## 5.1. Performance Under Different Load Levels

The data obtained from the conducted simulations demonstrate the efficiency of the proposed DRR-MDPF method under various traffic load conditions. In these experiments, metrics such as average throughput, Interest Satisfaction Ratio (ISR) at the node level, packet drop rate, average content retrieval time, and load balancing index were evaluated across scenarios with diverse background traffic loads and varying cache sizes.

Average throughput is defined as one of the key performance indicators of DRR-MDPF, representing the number of packets successfully delivered to their destination per unit of time [29]. The Interest Satisfaction Ratio (ISR) indicates the ratio of the number of data packets received to the number of Interest packets sent by all consumers [30, 31]. Additionally, the packet drop rate reflects the average number of packets lost within the network over a specified time interval [32]. The average delivery time shows the average duration required for each request to receive the desired data and serves as another important metric in the simulations [33, 34].

Finally, to measure load balancing among the network nodes, the coefficient of variation *(CoV)* index was used, defined as $CoV(f) = stdev([f(V)]) / E[f(V)]$, where $f(V)$ denotes the number of requests sent by node *V* [34].

Figure 3(a) shows the number of data packets received per second at the consumer node under varying Interest arrival rates from 2000 to 4000. This figure demonstrates that in all simulation scenarios, the proposed DRR-MDPF algorithm outperforms the other compared algorithms in terms of average throughput, regardless of the traffic load level. The DRR-MDPF algorithm exhibits a significant advantage over LA-MDPF, RFA, SAF, and SMDP at higher Interest rates, when the network experiences greater congestion. Specifically, at a rate of 4000 packets per second, the throughput value for DRR-MDPF is approximately 460.9, which is 5.6%, 12.6%, 18.9%, and 26.1% higher than LA-MDPF, RFA, SAF, and SMDP, respectively. The MDPF (Markov Decision Process Forwarding) module calculates the reward value of each interface based on key network metrics, and then the DRR module dynamically and adaptively allocates traffic shares, continuously favoring more efficient paths. This combination enables the proposed algorithm to make adaptive and optimal decisions, directing traffic through paths with the highest likelihood of success and capacity. In this algorithm, after receiving successful or unsuccessful feedback from each path, the share of that path is updated in the next decision round. This mechanism allows the system to quickly adapt to changes in network conditions and provide better performance. As network load increases, DRR-MDPF allocates a larger portion of traffic to more efficient interfaces, thereby utilizing network resources more effectively. However, under full router saturation, the improvement becomes limited because processing resources and queues in this state lack the capacity to handle additional requests. The reason for the bandwidth improvement in DRR-MDPF compared to RFA is that, instead of trying to evenly distribute the load across all available paths, DRR-MDPF focuses on paths with higher efficiency and dynamically manages their capacity based on the network's condition. This results in transmitting a greater number of packets per unit time, thereby providing higher bandwidth (12.6% more at the rate of 4000 packets per second) with only a slight reduction in load balancing. This performance difference becomes even more pronounced under heavy network load conditions.

Figure 3(b) shows the impact of cache size (ranging from 1% to 60%) on the average throughput of different algorithms. As expected, increasing the cache capacity improves the performance of all algorithms. However, the DRR-MDPF algorithm consistently outperforms the others across all cache sizes. At a cache size of 1%, the difference in average throughput between DRR-MDPF and the other algorithms is more pronounced; DRR-MDPF performs about 22% better than LA-MDPF and over 35% better than RFA. This highlights the high capability of DRR-MDPF in effectively utilizing limited network resources through intelligent scheduling and adaptive decision-making. For other cache sizes (15%, 30%, 45%, and 60%), DRR-MDPF continues to deliver the highest average throughput, although the performance gap between the algorithms decreases as cache size increases. Notably, around the 45% threshold, the growth rate of DRR-MDPF's throughput increases significantly again, which may be due to its ability to discover temporary content and make smart use of intermediate memory. At 60% cache size, all algorithms reach relatively high performance levels, but DRR-MDPF still maintains the highest throughput. This indicates that alongside its strong performance under limited resource conditions, the proposed algorithm also remains highly competitive when resources are abundant.

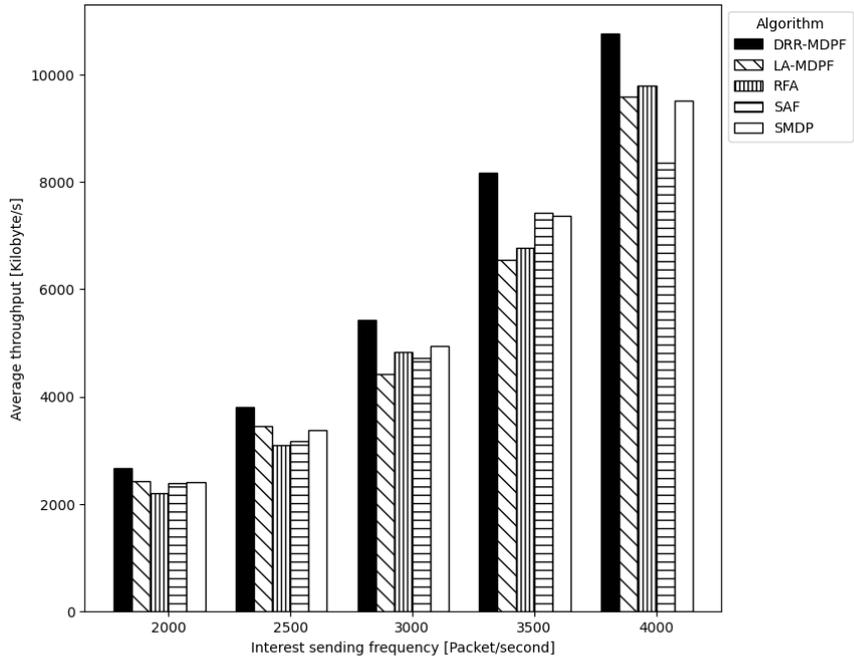

(a)

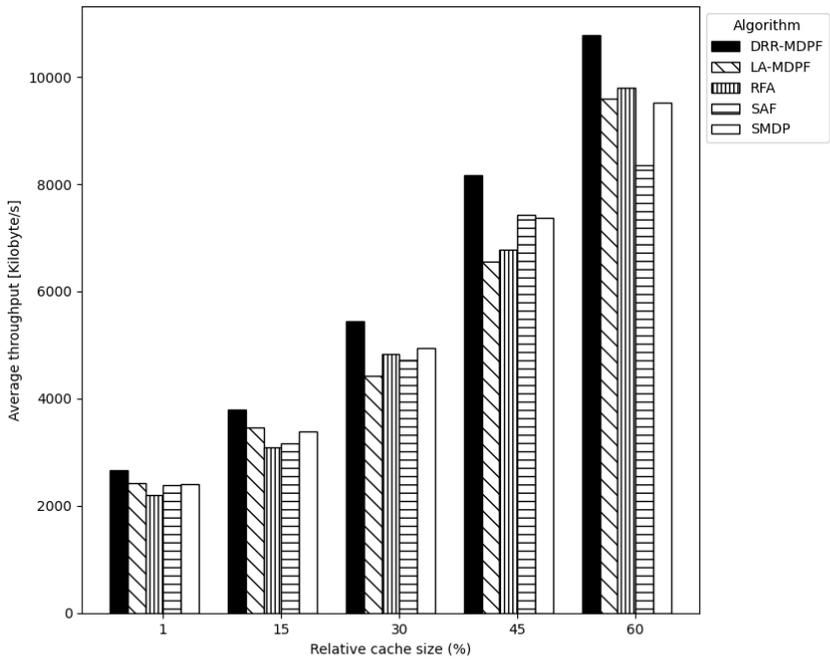

(b)

Figure 3. Average throughput under different conditions. (a) Cache size equal to 10% of total content. (b) Impact of increasing cache size under heavy traffic conditions.

Figure 4(a) illustrates the Interest satisfaction rate, measured in packets per second, for the consumer node as the Interest arrival rate varies from 2000 to 4000 packets per second. The data demonstrate that the proposed DRR-MDPF algorithm consistently outperforms competing methods such as LA-MDPF, SMDP, RFA, and SAF throughout the entire range of request rates. This advantage becomes even more pronounced under high traffic conditions when the network faces congestion and saturation. This enhanced performance stems from the integration of the MDPF-based intelligent decision-making mechanism with the DRR algorithm's fair and adaptive scheduling. Together, they facilitate the selection of paths that offer better responsiveness and lower latency. At the highest tested rate of 4000 requests per second, DRR-MDPF delivers an Interest satisfaction rate roughly 29% greater than SMDP and approximately 18% higher than LA-MDPF. Additionally, unlike some other method, this algorithm maintains robust performance under heavy traffic, highlighting its stability and ability to adapt to fluctuating network environments. The interface selection process in DRR-MDPF incorporates not only the history of prior outcomes but also adjustable probabilities, which helps avoid entrapment in local optima and promotes exploration of alternative routing options over time. This adaptability enables more efficient utilization of the network's resources and results in higher overall satisfaction of Interest requests within the NDN architecture.

Figure 4(b) illustrates the impact of cache size (ranging from 1% to 60%) on the Interest satisfaction rate. As expected, increasing cache capacity improves the performance of all strategies. However, the DRR-MDPF algorithm consistently outperforms the other methods across all cache sizes, with a more pronounced advantage at lower cache capacities, where resources are limited. At just 1% cache size, DRR-MDPF achieves a significantly higher satisfaction rate compared to other algorithms—about 22% better than LA-MDPF and over 35% better than RFA. This highlights DRR-MDPF's ability to maximize limited network resources through adaptive routing and intelligent scheduling. As the cache size grows up to 45%, the proposed algorithm demonstrates a faster increase in satisfaction rate relative to competing methods. This improvement is attributed to DRR-MDPF's superior capability to discover and leverage temporary content resources within the network. Beyond the 45% threshold, the benefit of increasing cache size diminishes, as other factors such as link bandwidth limitations and request patterns become the primary bottlenecks, limiting further performance gains.

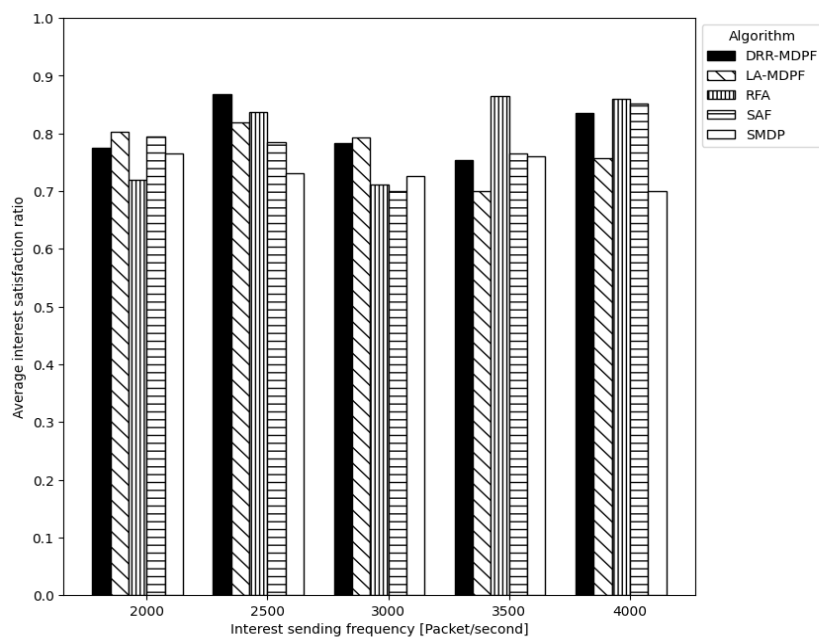

(a)

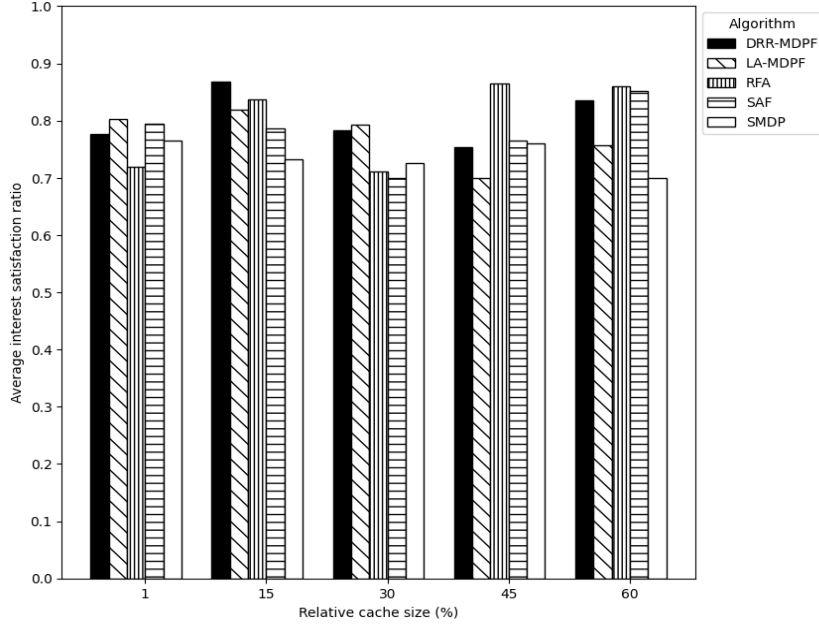

(b)

Figure 4. Average Interest satisfaction ratio at the node level. (a) Cache size at 10% of the content catalog. (b) Impact of increasing cache size under heavy traffic conditions.

Figure 5(a) illustrates the packet drop rate under varying traffic loads from 2000 to 4000 requests per second. As shown in the chart, the DRR-MDPF algorithm demonstrates acceptable and stable performance against increasing traffic rates, maintaining performance close to LA-MDPF up to a rate of 3500. However, at 4000 requests per second, when the network approaches full congestion, the packet drop rate for DRR-MDPF slightly exceeds that of LA-MDPF. This can be attributed to the DRR scheduling policy; under highly congested conditions, the algorithm may still allocate some traffic to lower-capacity paths to maintain overall network balance. In contrast, LA-MDPF, by focusing more on paths with higher rewards, manages to achieve a lower drop rate in critical situations. Nonetheless, the packet drop rate in DRR-MDPF remains lower than other compared methods such as RFA, BR, and SMDP, indicating its relative stability and effectiveness in managing network traffic even under heavy load. Overall, despite the slight degradation at 4000 requests per second, DRR-MDPF continues to deliver one of the best performances in maintaining network quality of service.

Figure 5(b) shows the impact of cache size, ranging from 1% to 60%, on the Packet Drop Rate for different routing strategies. In this chart, the proposed DRR-MDPF algorithm also demonstrates a more stable and acceptable performance compared to many other methods. At very limited cache sizes (such as 1% and 5%), all algorithms experience performance degradation, but DRR-MDPF still maintains a lower drop rate than RFA, BR, and SMDP. This indicates that DRR-MDPF can better manage packets and prevent their loss even under constrained resource conditions. As cache capacity increases to about 30%, the packet drop rate significantly decreases, and DRR-MDPF performs close to LA-MDPF and SAF. However, at larger cache sizes (for example, 45% and above), methods like LA-MDPF register slightly better performance compared to DRR-MDPF. This difference can be attributed to LA-MDPF's more direct use of reward data and its focus on paths with the highest cached content, while DRR-MDPF, with its DRR structure, still tries to balance traffic among different paths, which in scenarios with very rich caches may lead to less efficient cache utilization. Overall, DRR-

MDPF delivers competitive performance across all cache sizes and, in some cases, outperforms many other methods, especially under conditions with more limited network resources.

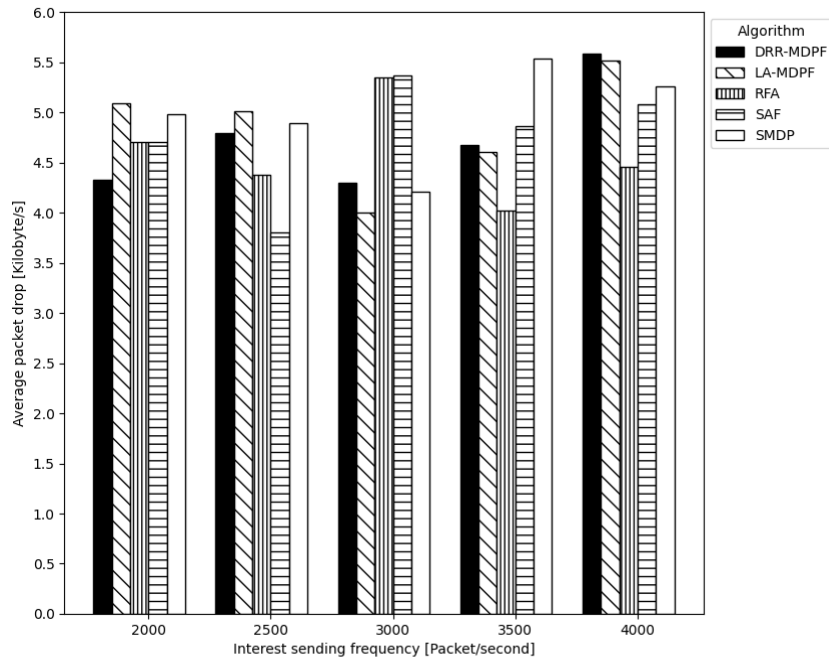

(a)

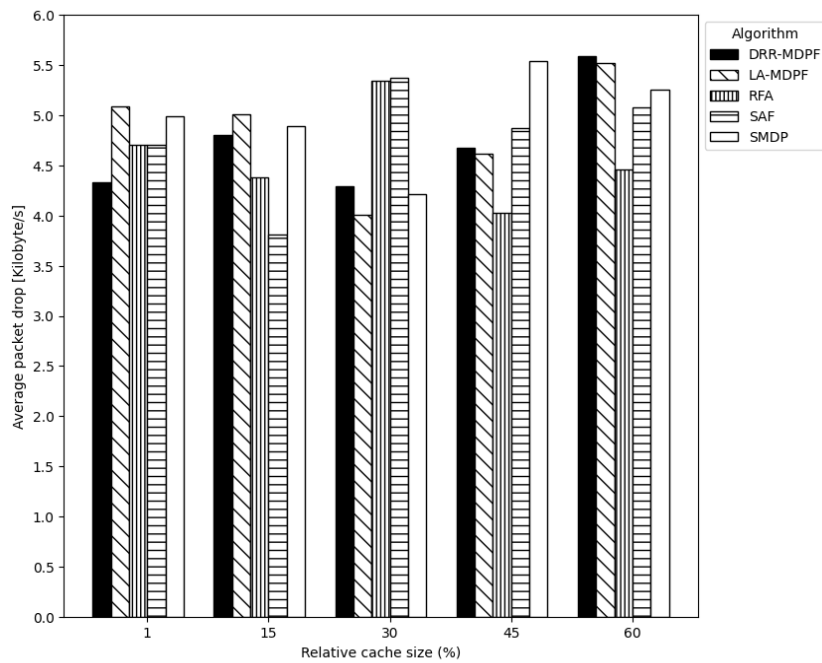

(b)

Figure 5. Average packet drop rate. (a) Cache size equivalent to 10% of the content catalog. (b) Impact of increasing cache size under heavy traffic conditions.

Figure 6(a) illustrates the average data retrieval time under different network traffic conditions. This metric represents the time required to receive data after sending an Interest, which is directly influenced by the efficiency of the selected paths and the level of congestion in the network. The chart shows that the proposed DRR-MDPF algorithm achieves lower retrieval times compared to other algorithms such as LA-MDPF, SAF, RFA, and SMDP across most traffic load levels. At an Interest rate of 4000 packets per second, representing the highest network congestion level, the retrieval time for DRR-MDPF is approximately 0.046 seconds, about 8%, 13%, 17%, and 22% lower than LA-MDPF, SAF, RFA, and SMDP, respectively. This favorable performance stems from the hybrid structure of DRR-MDPF, which uses an MDP-based decision policy to evaluate interfaces based on bandwidth, latency, and failure rate, and then adaptively allocates more efficient routing resources through the DRR mechanism. Together, these modules enable the selection of paths with the lowest delay and highest success rates in responding to Interests. Furthermore, the rapid update of rewards after each decision and the algorithm's quick adaptation to changing network conditions allow DRR-MDPF to maintain more stable performance under varying traffic loads.

Figure 6(b) illustrates the average packet drop rate at different cache sizes. This metric represents the percentage of packets that were lost due to reasons such as congestion or incorrect path selection, failing to receive a response. As shown in the graph, the proposed DRR-MDPF algorithm outperforms other strategies, including LA-MDPF, SAF, SMDP, and RFA, across all cache sizes. This advantage is especially pronounced when the cache size is very small (e.g., 1%). In this case, DRR-MDPF records the lowest drop rate of about 0.021, which is approximately 30%, 43%, 49%, and over 60% better than LA-MDPF, SAF, RFA, and SMDP, respectively. The main reason for this difference is DRR-MDPF's use of multi-criteria decision-making based on rewards related to bandwidth, delay, and the rate of unanswered Interests. While other algorithms like SAF or RFA rely on a single metric (such as RTT or response rate) for decision-making, DRR-MDPF selects paths by analyzing a more comprehensive view of the network state, focusing on routes with a lower likelihood of packet loss. Furthermore, the DRR algorithm's adaptive resource allocation helps prevent over-saturation of paths, reducing the chances of long queues and packet drops. Although all strategies show some improvement as cache size increases, DRR-MDPF consistently maintains the lowest drop rate, demonstrating its stable performance under varying resource conditions.

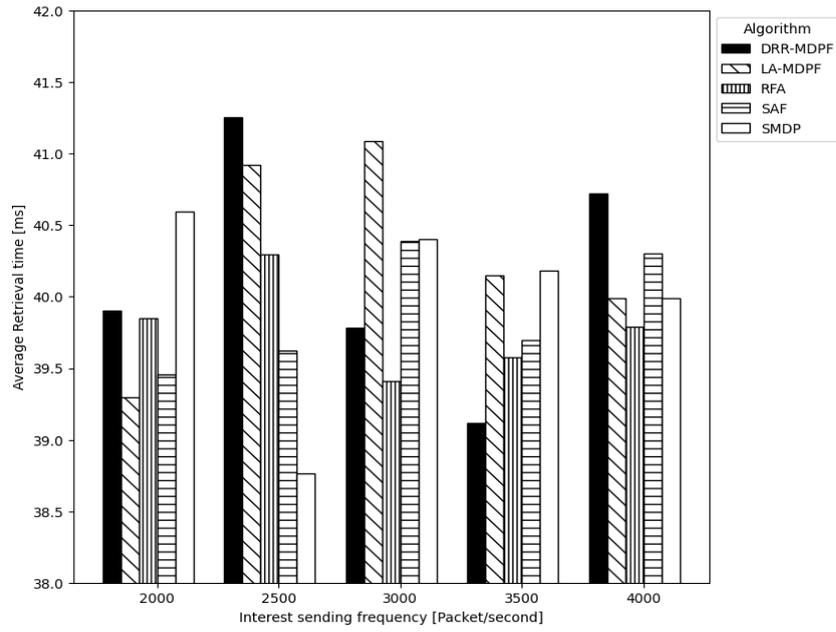

(a)

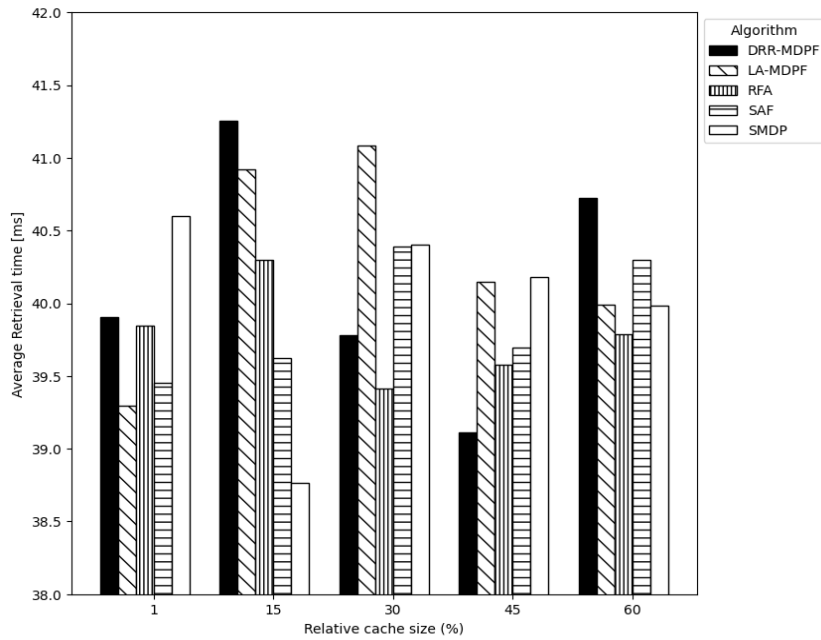

(b)

Figure 6. Average data retrieval time. (a) Cache size equivalent to 10% of the catalog. (b) Impact of increasing cache size under heavy traffic conditions.

Figure 7(a) illustrates the Load Balancing Factor for different algorithms under varying Interest sending rates. The results show that the RFA and SMDP algorithms maintain the highest levels of load balancing across all rates. This outcome aligns with the design philosophy of these algorithms, as they aim to distribute the network load uniformly across all available paths without considering adaptive factors related to the quality of the paths. In contrast, the proposed DRR-MDPF algorithm, despite registering lower Load Balancing Factor values compared to some reference methods, especially at high traffic rates (e.g., 4000 packets per second)—follows an intelligent and efficient load allocation approach. Instead of focusing solely on uniform traffic distribution, this algorithm combines adaptive learning based on MDP with dynamic scheduling to direct Interests toward paths that perform better in terms of Quality of Service (QoS) metrics such as bandwidth, delay, and success rate. Therefore, the relative decrease in the Load Balancing Factor for DRR-MDPF is not an indication of inefficiency but rather reflects a deliberate preference for high-performing paths to improve overall metrics like throughput and Interest satisfaction rate, as clearly observed in other evaluations (e.g., Figures 3 and 4). This is particularly important under heavy network load conditions, where optimal utilization of limited resources becomes critical.

Figure 7(b) compares the effect of relative cache size on the Load Balancing Factor for different algorithms. As observed, the SMDP and SAF algorithms maintain higher levels of load balancing across all cache sizes compared to other methods. This phenomenon primarily stems from the design nature of these algorithms, which focus on uniformly distributing traffic across available paths and avoid policies based on quality metrics. In contrast, the proposed DRR-MDPF algorithm, although exhibiting lower load balancing compared to some competitors, adopts an intelligent approach to managing Interests. By combining dynamic scheduling with a Markov decision process model, DRR-MDPF aims to route packets through paths that are more favorable in terms of metrics like bandwidth, delay, and dissatisfaction rate. Notably, as the cache size increases, traditional algorithms such as SMDP and SAF experience a significant rise in the Load Balancing Factor, whereas DRR-MDPF shows less fluctuation in this metric. This indicates that the proposed algorithm, relying on multi-criteria decision-making, is less dependent on cache size and maintains stable performance even under changing network conditions. Overall, despite having a relatively lower Load Balancing Factor, DRR-MDPF manages network resources more efficiently through its adaptive and informed design, prioritizing paths with higher quality of service, even if this comes at the expense of less uniform load distribution.

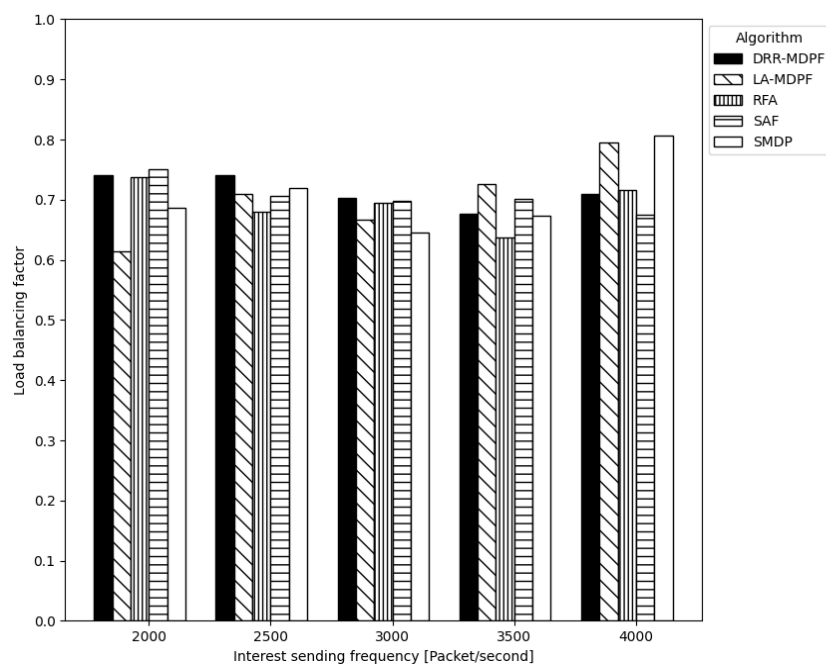

(a)

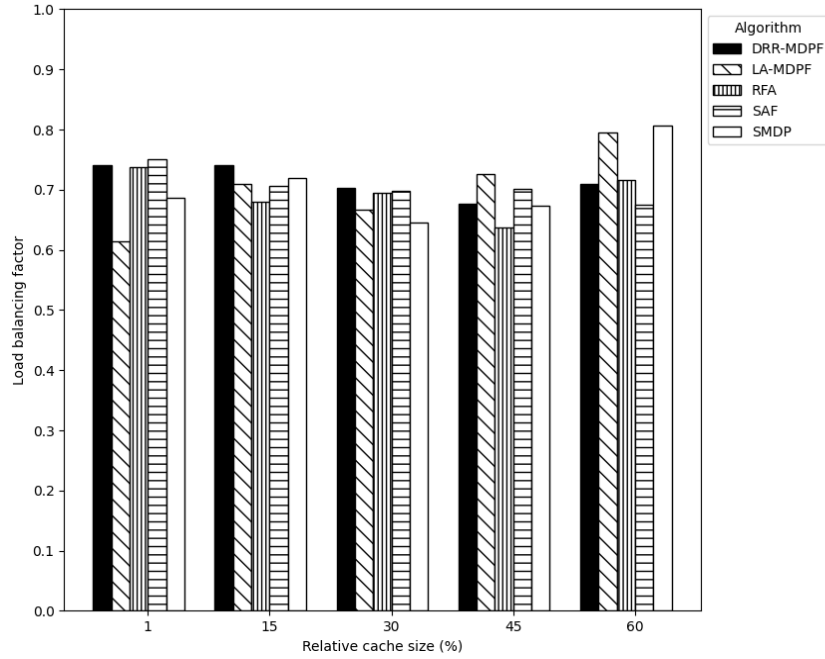

(b)

Figure 7. Load Balancing Factor in the network. (a) Cache size equal to 10% of the content catalog; (b) Impact of increasing cache size under heavy traffic conditions.

## 7. Conclusion

In Named Data Networking (NDN) architecture, designing a queue management and request forwarding strategy capable of adapting to the dynamic conditions of the network is of great importance. In this paper, a hybrid method called DRR-MDPF is introduced, which combines the Deficit Round Robin (DRR) algorithm and the Markov Decision Process Forwarding (MDPF) model to optimize resource allocation, reduce latency, and increase the overall efficiency of the network. In DRR-MDPF, the MDPF model evaluates the reward of each interface using three key metrics, available bandwidth, delay, and the rate of unsatisfied requests, while DRR acts as a fair mechanism to distribute traffic among the selected paths. This method leads to an intelligent and adaptive selection of routes that are optimized in terms of performance, stability, and Quality of Service (QoS). Simulations conducted in the ndnSIM framework under various traffic scenarios, cache sizes, and connectivity levels showed that DRR-MDPF outperforms reference algorithms such as SAF, RFA, SMDPF, and LA-MDPF in metrics including Interest Satisfaction Rate (ISR), data retrieval time, packet drop rate, throughput, and load balancing. Additionally, due to its single-path structure, despite targeted utilization of optimal paths, DRR-MDPF is less complex computationally compared to multipath methods and demonstrates high stability in resource-constrained environments. Another important feature of DRR-MDPF is its high adaptability to topology changes and traffic load variations, achieved through adaptive learning and continuous feedback. Ultimately, it can be concluded that DRR-MDPF, as an intelligent and forward-looking strategy, can ensure optimal and stable network performance when facing key challenges in NDN, especially in high-load scenarios and dynamic environments.

**Authorship contribution statement**

**Fatemeh Roshanzadeh:** Conception and design of study/review/case series, Acquisition of data, laboratory search, Drafting of article.
**Hamid Barati:** Conception and design of study/review/case series, Analysis and interpretation of data collected, Final approval and guarantor of manuscript.
**Ali Barati:** Acquisition of data, Laboratory or clinical/literature search, Analysis and interpretation of data collected, Final approval.

**Declaration of competing interest**
The authors declare that there is no conflict of interests regarding the publication of this manuscript.

**Data availability**
No data was used for the research described in the article.

**Acknowledgments**
None.

**Ethics approval**
This article does not contain any studies with human participants.

**Funding Declaration**
No Funding

**Clinical trial number:** not applicable